# Internet of Things (IoT) and Cloud Computing Enabled Disaster Management

**Raj Gaire[1], Chigulapalli Sriharsha[2], Deepak Puthal[3], Hendra Wijaya[4], Jongkil Kim[4], Prateeksha Keshari[5], Rajiv Ranjan[1,6], Rajkumar Buyya[7], Ratan K. Ghosh[2], RK Shyamasundar[5] and Surya Nepal[4]**

[1] CSIRO Data61, GPO Box 1700, Canberra ACT 2601, Australia
[2] CSE, IIT Kanpur, Kanpur 208016, India
[3] SEDE, University of Technology Sydney, PO Box 123, Broadway, NSW 2007, Australia
[4] CSIRO Data61, PO Box 76, Epping NSW 1710, Australia
[5] CSE, IIT Bombay, Powai Mumbai 400076, India
[6] SCS, Claremont Tower, Newcastle University, NE1 7RU, UK
[7] CIS, University of Melbourne, Parkville VIC 3010, Australia

**Abstract**

Disaster management demands a near real-time information dissemination so that the emergency services can be provided to the right people at the right time. Recent advances in information and communication technologies enable collection of real-time information from various sources. For example, sensors deployed in the fields collect data about the environment. Similarly, social networks like Twitter and Facebook can help to collect data from people in the disaster zone. On one hand, inadequate situation awareness in disasters has been identified as one of the primary factors in human errors with grave consequences such as loss of lives and destruction of critical infrastructure. On the other hand, the growing ubiquity of social media and mobile devices, and pervasive nature of the Internet-of-Things means that there are more sources of outbound traffic, which ultimately results in the creation of a data deluge, beginning shortly after the onset of disaster events, leading to the problem of information tsunami. In addition, security and privacy has crucial role to overcome the misuse of the system for either intrusions into data or overcome the misuse of the information that was meant for a specified purpose. These problems can be addressed by processing the collected data in real-time and extracting meaningful and actionable information for emergency providers while taking care of security and privacy aspects. Such situation-awareness applications demand a large amount of computing resources. The cloud, which provides elastic and scalable infrastructure, becomes the natural choice for such applications. In this chapter, we provide such a situation aware application to support disaster management data lifecycle, i.e. from data ingestion and processing to alert dissemination. We utilize cloud computing, Internet of Things and social computing technologies to achieve a scalable, efficient, and usable situation-aware application called Cloud4BigData.

## Introduction

A disaster can come in many forms including but not limited to earthquakes, hurricanes, foods, fire and outbreak of diseases. It causes loss of lives and severely affects the economy [1]. In the 2009, Victorian bushfires in Australia costed $4.3 billion and caused 173 human fatalities with over 1800 homes destroyed. The Hazelwood coalmine fire in 2014 [2], which costed over $100 million, had severe impact on the long term health of the affected people. The Emergency Events Database (EM-DAT) figures [3] indicate that there were 346 natural disaster events occurred globally in 2015 alone. Due to these events, a total of 98.6 million people were affected and 22,773 people died. The earthquake in Nepal cause 8,831 deaths, the most death from a single event in that year. The economic cost of these events was a massive US$66.5 billion.

According to United Nations International Strategy for Disaster Reduction (UN/ISDR) [4], a disaster is a serious disruption of the functioning of a community or a society, at any scale, frequency or onset, due to hazardous events leading to impacting human, material, economic and environmental losses. The source of disaster can be natural, anthropogenic or both. Natural disasters are associated with natural processes and phenomena such as hurricane, tsunami and earthquake, while anthropogenic disasters are predominantly induced by human activities, e.g. civil war. Since the occurrences of disasters cannot be completely eliminated, the effort is focused on the better management of disasters across different phases of disaster management life cycle [1], i.e. mitigation, preparedness, response and recovery. The main goal of a disaster risk management (DRM) strategy is to reduce the impact of disaster on human lives and economy. Information and communication technologies (ICTs) have already been used to support the DRM activities [5]. For example, computer modelling are used to forecast natural disaster warning such as the probability of flood and fire, and the path of a hurricane. Similarly, various communication technologies are used to disseminate information before, during and after the occurrence of a disaster event.

Inadequate situation awareness in disasters has been identified as one of the primary factors in human errors, with grave consequences such as deaths and loss of critical infrastructure. Timely acquisition and processing of data from different sources and extraction of accurate information plays an important role in coordinating disaster prevention and management. For instance, during the 2010 Queensland flooding in Australia, Queensland Police (QP) [6] analysed messages posted on social media by people in the affected regions to understand the situation on the ground and appropriately coordinate search and rescue operations. However, there is a pitfall. The growing ubiquity of social media and mobile devices, and pervasive nature of the Internet-of-Things means that there are more sources of data generation. Shortly after the onset of a disaster event, the volume of data dramatically increases which ultimately results in the creation of a tsunami of data. For example, during the 2010 Haiti earthquake [7], text messaging via mobile phones and Twitter made headlines as being crucial for disaster response, but only some 100,000 messages were actually processed by government agencies due to lack of automated and scalable ICT infrastructure. The number of messages has been continuously growing with over 20 million tweets posted during the Hurricane Sandy in 2012. This data tsunami phenomenon, also known as the BigData problem, is a new grand challenge in computing [8, 9].

Internet of things (IoT), cloud computing and big data are three disruptive technologies which have potential to make significant impact towards ad-dressing the above problem. These technologies are already creating impact in our everyday lives. The IoT systems including sensors, sensor networks and mobile phones have been used to monitor local environmental conditions. Today, farmers can afford to install their own weather stations and use the localised data to precisely predict the conditions of their farms rather than relying on the weather information of their region. Businesses have already started to exploit the value within the large volume of data [10]. For example, a supermarket knows what a consumer is going to purchase in his/her next shopping even before the consumer makes the decision. Similar impacts have been delivered by cloud computing. Organisations are already outsourcing their IT infrastructures to cloud vendors because of not only cost saving but also its offering of high scalability and availability. Big data analytics requires a large computing IT infrastructure. Building such a large computing infrastructure is impossible for some businesses while it can be very costly for others. Cloud computing helps to lift this burden by offering a highly available and pay-as-you-use IT infrastructure that can meet the demands in a foreseeable future. Cloud computing has also become a natural choice of computing infrastructure for big data analytics. As these technologies progress, IoT and cloud computing enabled big data analytics will become inevitable in development of novel disaster management applications [11, 12].

In this section, we first conceptualise a scenario where IoT, cloud computing and big data technologies work together to mitigate a flood disaster event. We then analyse the gaps among these technologies to develop our situation-awareness application framework in the following sections.

## *Motivating Scenario: Mitigating Flood Disaster*

In Australia, the Crisis Coordination Centre (CCC) is responsible for large-scale disaster management. The CCC is a round-the-clock all-hazards management facility that provides security, counter terrorism, and the monitoring and reporting of natural disasters and other emergencies. The CCC has policies and procedures for the tasks to be undertaken during disaster events. Part of its remit is the analysis of data from multiple sources to understand the scope and the impact of a disaster event. Figure 1 illustrates a flood disaster management scenario where CCC needs to manage the people living in the disaster area and reduce the impact on health and wellbeing of the individuals. Here we present an imaginary flood disaster scenario to illustrate how IoT and Cloud Computing enabled BigData technologies can help to mitigate the disaster event.

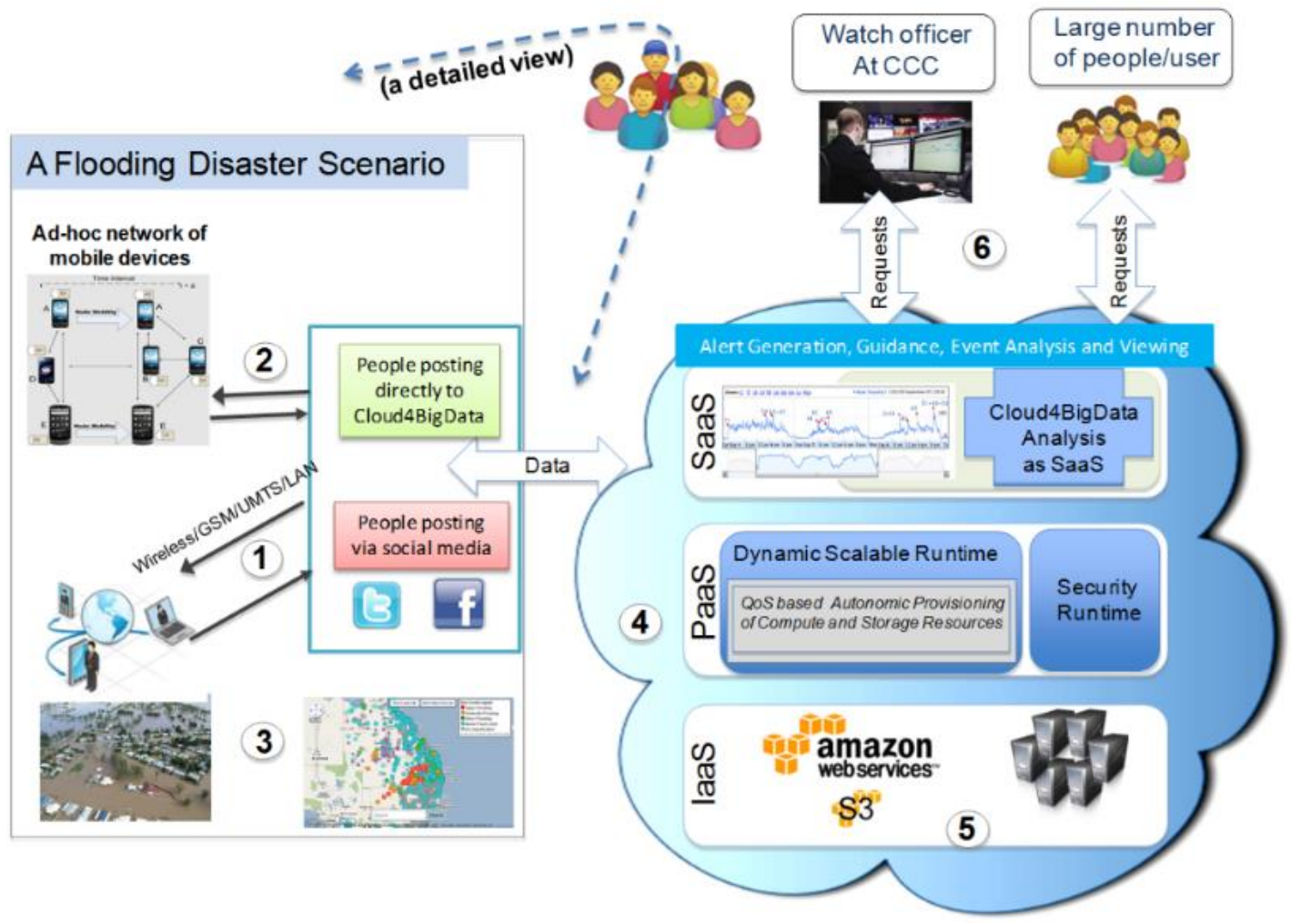

*Figure 1: A flood disaster management scenario*

Imagine that a weather forecast in a certain area indicates a heavy rainfall which can possibly cause flooding in the area. After receiving information from Bureau of Meteorology, the CCC creates a transient social network to provide targeted information to the people in the area. The telecommunication providers provide a list of landline phone numbers installed in the area, as well as a list of mobile phones that are roaming around in the area. This information is used to contact and encourage people in the area to use mobile app, register their apps to receive emergency situation information and enable ad-hoc networking when necessary. Besides the mobile app, hotlines and SMS services are provided for people to contact. Similarly the radio and television communication channels are used for mass dissemination of alerts. As expected, severe flooding has caused electricity outage in some areas. People in those areas are still able to collect information through their mobile devices. People are in touch with their families and friends using the transient network and social media such as Twitter and Facebook. Finally, sensors monitoring weather, water quality, air quality etc. are providing crucial information that can affect the health and wellbeing of people living in the area.

From the technical perspective, collection and integration of large volume and variety of information had been daunting and time consuming in the past. Such complex and time-consuming operations of real-time analysis of streaming data from multiple digital channels such as social media, mobile devices and SMS gateways have been offloaded to a software service provider. As illustrated in the figure, the SaaS platform has implemented the analysis algorithm and provision tools for calculating awareness of the situation on the ground. CCC coordinators are receiving this information on their computer screens. The information helps to make decisions based on the information at the instance.

### *The Problem*

In order to achieve the above scenario in reality, Cloud4BigData needs to function as a reliable repository and aggregator of data from various channels including sensors, social media and mobile applications at a higher level of granularity. It is also clear that there is an immediate need to leverage efficient and dynamically scalable ICT infrastructure to analyse BigData streams from various digital channels in a timely and scalable manner to establish accurate situation awareness [13] during disaster events. We need a complete ICT paradigm shift in order to support the development and delivery of big data applications. On one hand, we need to make use of IoT devices, delay tolerant networks (DTNs) [14] and transient social networks (TSNs) [15] to gather more data during a disaster. On the other hand, we need to deploy ICT infrastructure especially for disaster management, in a way that the downstream applications do not get overwhelmed by incoming data volume, data rate, data sources, and data types. In addition, such system should not compromise the security and privacy of the users. We propose to achieve this by, firstly, extending DTNs and TSNs for disaster management applications, secondly, leveraging cloud computing systems to engineer and host next-generation big data applications, and finally, deploying various security mechanisms.

## Background

This section will provide a brief introduction to the relevant technologies and their limitations in the context of a disaster management application.

### *Internet of Things (IoT)*

As defined by Minerva, Biru [16], an Internet of Things (IoT) is a network that connects uniquely identifiable *Things* to the Internet. The *Things* have sensing/actuation and potential programmability capabilities. Through the exploitation of unique identification and sensing, information about the *Things* can be collected and the state of the *Things* can be changed from anywhere and anytime. Here, we emphasis sensors, mobile phones and their networks as the key components of an IoT.

#### Sensors and Mobile Phones

Different terminologies have been used to describe sensors and sensor networks. The W3C semantic sensor network working group analysed these terminologies to develop SSN ontology [17] which states:

> "Sensors are physical objects that perform observations, i.e., they transform an incoming stimulus into another, often digital, representation. Sensors are not restricted to technical devices but also include humans as observers"

The low cost of electronic sensors has made them the technology of choice for collecting observation data. For example, in the context of disaster management, sensors have been used to measure weather conditions such as temperature, relative humidity and wind direction/velocity. Among various usages, these measurements have been used to forecast the fire danger ratings of bushfires in Australia [18], as well as, to assess and plan for the fire-fighting activities during bushfires. Similarly, with the ubiquity of smartphones, people are now at the forefront of generating observation data. Again in the context of the disaster management, people have been using social media to publish disaster related observations. These observations can be used for early detection of a disaster [19], as well as, for the situation awareness during the disaster event [20].

Traditional IoT and smartphone based applications often assume that a communication network exists between an IoT device and the internet. This assumption does not always hold, particularly in the following three situations. First, the cellular network often does not exist in unpopulated remote areas. Moreover in some developing countries, the cellular network may not exist at all even in well populated areas. Second, physical infrastructures including the telecommunication infrastructures can be severely damaged during catastrophic disasters. Moreover, the electricity outage making the telecommunication network inaccessible can also be widespread [21]. Third, the communication system can be severely disrupted due to dramatic increase in the demand of services during a disaster situation making the system temporarily unavailable. In order to prepare for these situations, alternative approaches need to be developed.

### Delay Tolerant Networks

Delay Tolerant Networks (DTNs) provide alternatives to traditional networks. A DTN addresses three distinct problems related to communication in challenged networks: delay, disruption and disconnection. The class of networks that belong to DTN are: inter planetary networks (IPN), mobile ad hoc networks (MANET), vehicular ad hoc networks (VANET), mule networks and wireless sensor networks (WSN). A DTN facilitates connectivity of systems and network regions with sporadic or unstable communication links. The connections among DTN based systems exist between two extremities, from strong intermittent connections to full disconnection. From a user's perspective, a DTN is an opportunistic network which provides a possibility of reliable communication in situations when the networks are severely disrupted and characteristically unreliable. In a severe disaster situation, the communication may be challenged and should be channelized opportunistically over these networks.

When creating this type of opportunistic network, the three dimensions, namely, technological, protocol and mobility need to be considered:

*Technological dimension*: An opportunistic network is primarily based on wireless links. The characteristics of these links may vary significantly among devices. For example, some links could be established using Wi-Fi technology while others may use Bluetooth or even proprietary technologies. The differences in these technologies influence encoding, modulation, error correction code and latency among other physical features related to communication.

*Protocol dimension*: Similar to technology, protocols also vary significantly with variations in communication technologies. MAC protocols, network protocols, and transport protocols are dependent on the network standards followed by physical links. For example, the format of MAC frames for Bluetooth based on IEEE 802.16 is very different from that of Wi-Fi which is based on IEEE 802.11. Due to self-organising nature of a DTN, the routing protocols are required to be self-adaptable.

*Mobility dimension*: Considering most of the end devices during severe disasters are portable and mobile devices, the mobility dimension is critical to connectivity, routing and coverage. Since the devices are carried by people, data can be carried forward physically, albeit slowly, closer to the destination. Furthermore, the connection may be intermittent due to the coverage problem. The user of the device may also opt for a voluntary disconnection to save battery power.

Some of the challenges in dealing with DTNs are [14]:

*Intermittent connectivity*: The end-to-end connection between communicating systems may not exist.

*Low data rate and long latency*: In DTNs, transmission rates are comparatively low and latency may be large.

*Long queuing delay*: The queuing delay is the time taken to flush the earlier messages out of the queue. This could be long in DTNs.

*Resource scarcity*: The energy resources of nodes in DTNs are often limited. This could be because of their mobility or destruction of infrastructure in that area.

*Limited longevity*: The links could be intermittent and may not last long.

*Security*: Because intermediate nodes can be used to relay messages, there are possibilities of security attacks that can compromise information integrity, authenticity, user privacy and system performance.

The fundamental problem in DTN is routing of messages when the receiver and the sender are not connected by a network at the time of dispatch. Storing and forwarding transmission is the only way of communication in a challenged network which must also be able to sustain delays, disruption and disconnection.

### Transient Social Network

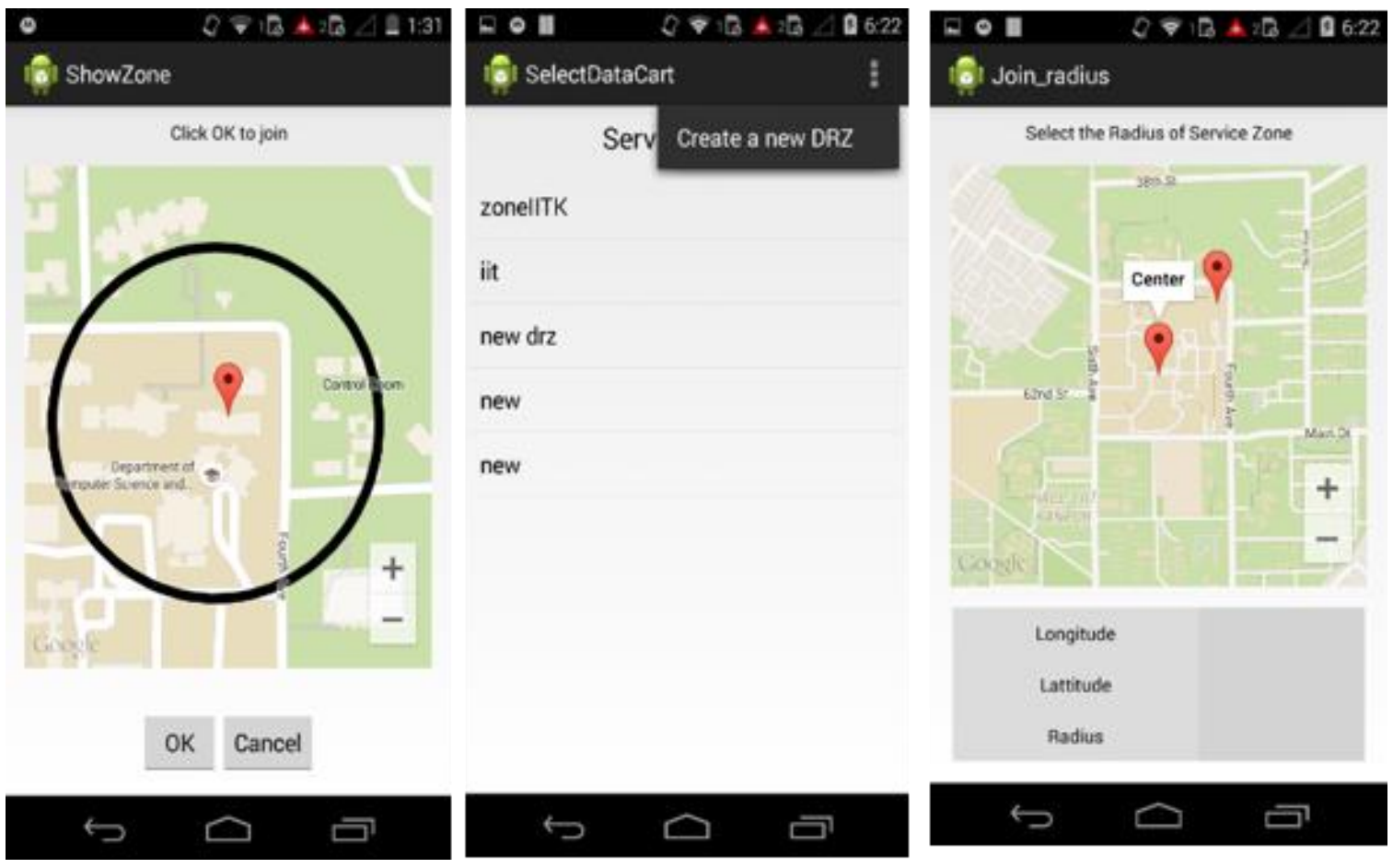

*Figure 2: Creating and joining a centre of activity with a defined radius of activity in TSN*

Transient social network (TSN) has been evolved as a mobile, peer-to-peer on-demand community inspired network. For example, a TSN can be created by using applications through which a person, in distress, can send messages to others who have also deployed the TSN application in order to seek support [15]. More formally, *a TSN is a spatio-temporal network of people having common interests*. A TSN application can be developed over the existing social networks such as Facebook and Twitter, as well as by creating new dedicated applications for smartphones. In case of an emergency situation, a TSN can be created over DTN using mobile phones to communicate between the nodes when the telecommunication network is unavailable during the time of the disaster [15]. The use case scenario of a TSN most closely resembles the type of publish-subscribe system proposed called Kyra [22]. The Kyra approach combines both filter-based and event-based routings to create a brokered network architecture that captures spatio-temporal importance in publishing information. However, Kyra has three important limitations: (I) spatial locality is based only on network proximity, (ii) participating nodes are static and homogeneous, and (iii) it does not admit priority on message dissemination. A TSN not only provides flexibilities in all the three limitations of Kyra but also exists opportunistically for fulfilment of certain community centred activities including disaster management as illustrated in Figure 2.

**TSN over DTN**

In disaster situations where some network infrastructure is destroyed by calamities such as flood or earthquake, the active communication network can be viewed as a collection of disconnected islands of network partitions. As a part of the framework, we conceptualise TSN over DTNs through a mule transport layer which provides a workable and practical solution for communication over challenged networks. Specifically, a TSN application needs to be established along with DTN using a single-hop communications in an immediate neighbourhood that is devoid of any network infrastructure. The messages created at source devices are transported to other network partitions through the *mobile mules* which use DTN features like bundling. As such, our TSN network contains nodes which act as message generators, message collector, message distributor and mule. These roles are described as below.

*Message Generator:* These are nodes which want to send messages across. The message destination could be inside the same island or in another disconnected island. A distress node can be termed as a message generator.

*Message Collector:* Since the Internet or cellular network is unavailable, we need a node that can collect the messages until the messages can get delivered to the other islands. This, here, is done by a local mule node which bundles the messages sent by the generators. The collector also receives messages from the distributor and then disseminates the message to their respective recipients. In case the cellular network or DTN is working, the collector will also send the message using that network. There can be an auxiliary collector which will be functional as message collector when an existing message collector goes offline.

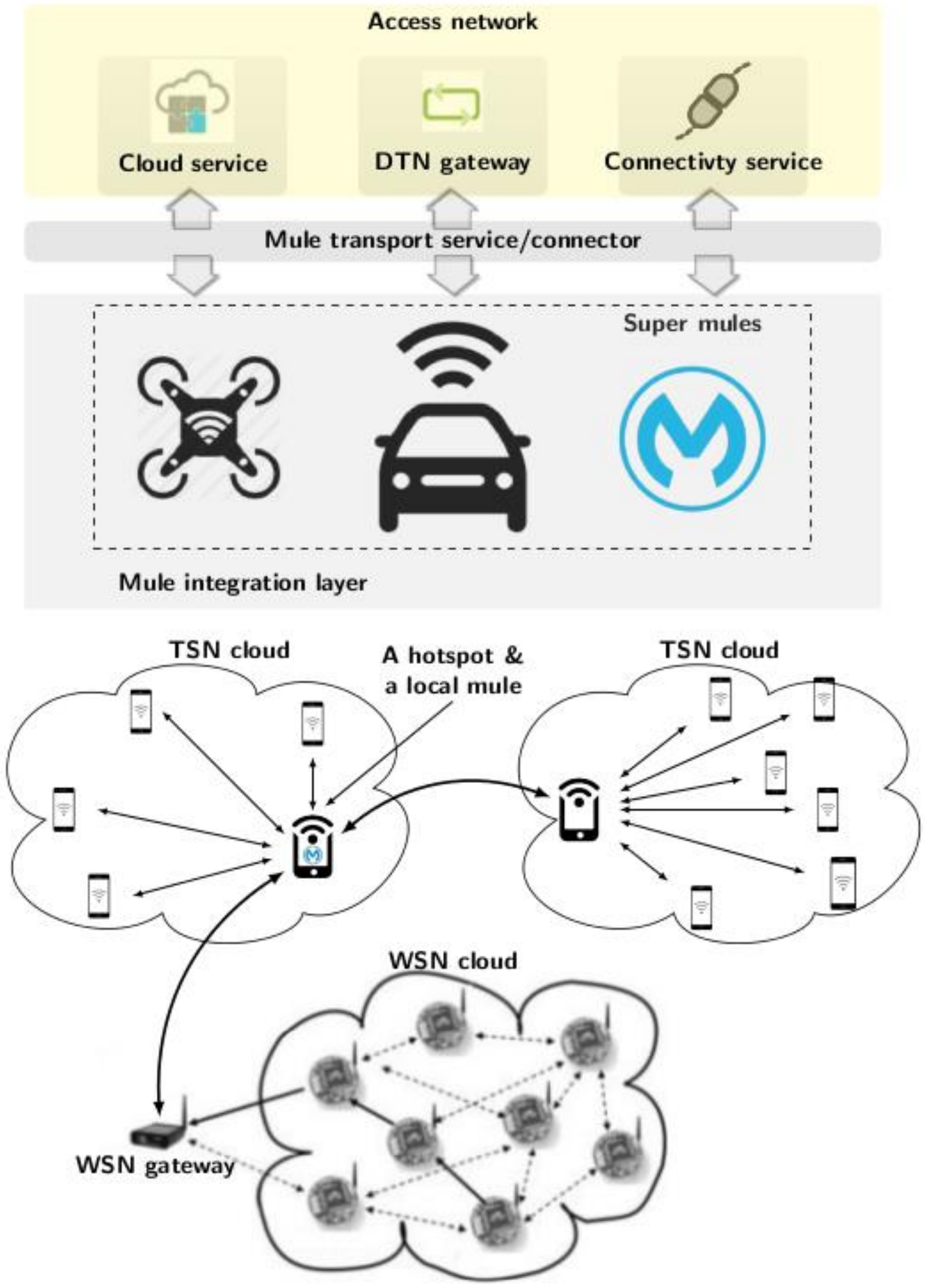

*Figure 3: A framework of Delay Tolerant Network in disaster management*

*Message Distributor:* This role is assumed by the super mules which move between the disconnected islands. A collector will transfer the message bundle to a distributor whenever a distributor comes in range of communication (wi-fi range). The distributor then receives all the messages of an island from the collector and then moves into the range of another collector of another island. This second collector will receive the messages from the distributor, unbundle the messages since it has received messages as a bundle. After unbundling, the collector disseminates the message to the respective target.

*Mules:* They collects messages from across the island and dumps them to the message collector.

In a geographically connected island, TSNs can be viewed as micro-islands and connect the network partitions of the island through local mule service. In such islands, smartphone based nodes establish DTN by enabling mobile app based Wi-Fi hotspot. The node acting as a hotspot also provides the services of a Message Collector handling the bundling functions for TSN. In addition, dedicated message collectors in an area may also be introduced as a part of the disaster risk management action. Thus, the nodes running TSN application may additionally provide local mule service as indicated in Figure 3. All the distress nodes can connect and transfer the messages to the mule. Some nodes can be mounted on drones or other autonomous agents travelling between the islands. Whenever these nodes, referred to as super mules, come into contact with a Message Collector, an exchange of message bundles takes place. The local mule can also receive the message bundled from the super mule, then spread the message to the recipients.

In general, the TSN is not bound by the range of the Wi-Fi. In fact, the TSN is defined by loosely connected nodes that can interact with each other using Wi-Fi or other communication protocols. Since the nodes can move around, the geographical range of TSN is not limited by the range of Wi-Fi hotspot span of the central node. Moreover, the same node can also carry on different roles of a TSN. Also, instead of a mobile super mule, it is possible to establish a server application which may leverage Internet facilities provisioned through High Altitude Aeronautical Platform Stations (HAAPS) for radio relay capabilities [23, 24]. Google's net-beaming Balloon or Facebook's laser Drones based on the HAAPS concept are still being implemented. The proposed opportunistic network architecture can easily be adapted to HAAPS radio relay capabilities as and when they become a reality. When a Message Collector (may be on move) comes into range of internet, it uploads the bundles to the server application. The bundles then can be distributed to respective destination island's Message Collectors for final distribution to end hosts again by running TSN cloud application.

## *Cloud Computing*

Cloud computing is the fulfilment of the vision of Internet pioneer Leonard Kleinrock [25] who said in 1969:

> "Computer networks are still in their infancy, but as they grow up and become sophisticated, we will probably see the spread of 'computer utilities' which, like present electric and telephone utilities, will service individual homes and offices across the country."

Today, cloud computing [26, 27] assembles large networks of virtualised ICT services such as hardware resources (e.g. CPU, storage, and network), software resources (e.g. databases, application servers, and web servers) and applications. Cloud computing services are hosted in large data centres, often referred to as data farms, operated by companies such as Amazon, Apple, Google, and Microsoft. The term cloud computing has been used mainly as a marketing term in a variety of contexts to represent many different ideas [28]. This has resulted in a fair amount of scepticism and confusion. In this chapter, we use the definition of cloud computing provided by the National Institute of Standards and Technology (NIST) [29] as:

> "Cloud computing is a model for enabling convenient, on-demand network access to a shared pool of configurable computing resources (e.g., networks, servers, storage, applications, and services) that can be rapidly provisioned and released with minimal management effort or service provider interaction".

### Essential Characteristics

Cloud computing provides unique characteristics that are different from traditional approaches. These characteristics arise from the objectives of using clouds seamlessly and transparently. The five key characteristics of cloud computing defined by NIST [29] include:

*On-demand self-service:* users can request and manage resources such as storage or processing power automatically without human intervention.

*Ubiquitous network access:* computing resources can be delivered over the Internet and used by a variety of applications.

*Resource pooling:* users can draw computing resources from a resource pool. As a result, the physical resources become 'invisible' to consumers.

*Rapid elasticity:* consumers can scale up and down the resources based on their needs.

*Measured Service:* often called as Pay As You Go (PAYG), offers computing resources as a utility which users pay for on a consumption basis.

**Service Models**

In addition to the above five essential characteristics, the cloud community has extensively used the following three service models to categorise the cloud services:

*Software as a Service (SaaS)*. Cloud providers deliver applications hosted on the cloud infrastructure as internet-based on-demand services for end users, without requiring them to install these applications. Examples of SaaS include SalesForce.com and Google Apps such as Google Mail and Google Docs.

*Platform as a Service (PaaS)*. PaaS provides a programming environment such as OS, hardware and network so that users can install software or develop their own applications. Examples of PaaS include Google AppEngine, Microsoft Azure, and Manjrasoft Aneka.

*Infrastructure as a Service (IaaS)*. IaaS provides a set of virtualised infrastructural components such as processing units, networks and storage. Users can build and run their own OS, software and applications. Examples of IaaS include Amazon EC2, Azure IaaS, and Google Compute Engine (GCE).

These three services sit on top of one another, IaaS at the bottom and SaaS at the top, and form the layers of cloud computing.

**Deployment Models**

More recently, four cloud deployment models have been defined to deliver cloud services to users:

*Private cloud*. In this model, computing resources are used and controlled by a private enterprise. The access to resources is limited to the users that belong to the enterprise.

*Public cloud*. In this model, computing resources are dynamically provisioned over the Internet via web applications/web services.

*Community cloud*. In this model, a number of organizations with similar interests and requirements share the cloud infrastructure. The cloud infrastructure could be hosted by a third-party vendor or within one of the organizations in the community.

*Hybrid cloud*. In this model, the cloud infrastructure is a combination of two or more clouds (private, community or public) described above. This model is becoming popular as it enables organisations to increase their core competencies by outsourcing peripheral business functions onto the public cloud while controlling core activities on-premises through a private cloud.

While cloud computing optimises the use of resources, it does not (yet) provide an effective solution for hosting disaster management related big data applications to analyse tsunami of data in real time [30] due to multiple reasons. First, most of the researches in the cloud computing space have been devoted to managing generic web-based applications, which are fundamentally different from big data applications. Second, current cloud resource programming abstractions are not at the level required to facilitate big data application development and cloud resource provisioning, defined as the process of selection, deployment, monitoring, and runtime management of hardware and software resources to ensure QoS targets of big data applications. Finally, the system should be secure, i.e., the privacy of the citizens is maintained, the integrity of the data exchanged is guaranteed and the service is always available.

## ***Big Data***

Big data computing is an emerging data science paradigm of multidimensional information mining for scientific discovery and business analytics over large-scale infrastructure [10]. There is not a single standard definition of big data. Industry, academic and standard communities are defining big data from their own perspective. The following definitions from Gartner and National Institute of Standards and Technology (NIST) reflect the concept of big data from different angles.

Gartner first defined big data as the three Vs: Volume, Velocity, and Veracity [31]. This is the most well-known and widely-accepted definition of big data in scientific and business communities. Gartner now formally defines big data as "Big data is high-volume, high-velocity and high-variety information assets that demand cost-effective, innovative forms of information processing for enhanced insight and decision making" [32].

NIST offers the following definition [33]:

> "Big Data refers to the inability of traditional data architectures to efficiently handle the new datasets. Characteristics of Big Data that force new architectures are volume (i.e., the size of the dataset) and variety (i.e., data from multiple repositories, domains, or types), and the data in motion characteristics of velocity (i.e., rate of flow) and variability (i.e., the change in other characteristics)".

**Big Data Lifecycle**

NIST defined [33] four stages for big data lifecycle are shown in Figure 4**Error! Reference source not found.**.

*Collection*: At this stage, the data is collected from a variety of sources where the data is generated. The outcome of this process is a collection of raw data. There are a number of significant challenges in this phase such as data ingestion, transportation, and storing.

*Preparation*: This stage processes the collected raw data to remove corrupt and inaccurate data and produce the quality data. The significant challenges at this stage are to check the veracity, integrity and authenticity of the data.

*Analysis*: The analysis phase takes the cleansed information and generates knowledge that can be used to take actions. The challenges at this stage are to develop efficient and effective algorithms to process large volume of data including resource allocation and provisioning.

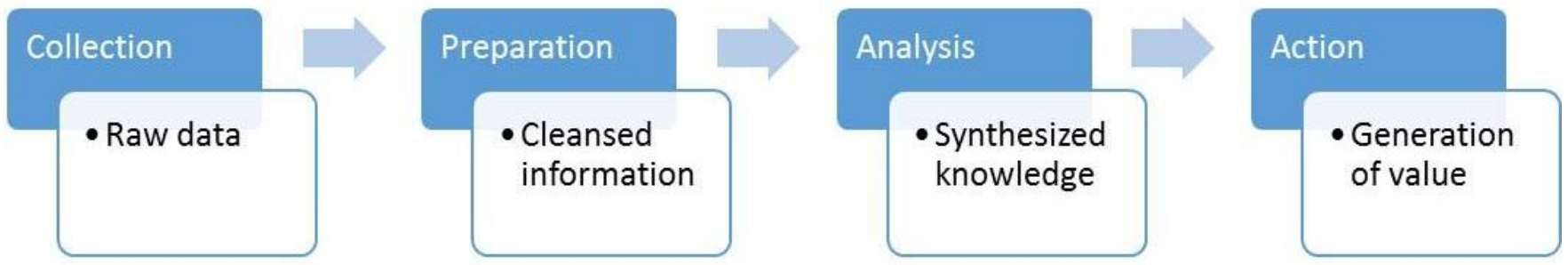

*Figure 4: Big Data Life Cycle*

*Action*: The last phase of the big data life cycle is to take necessary action or make decisions based on the knowledge extracted from the analysis phase [34]. The actions depend on the application, for example dispatching paramedic services to an identified location in case of an identified medical emergency.

Over the years big data technologies have been evolved from providing basic infrastructure for data processing and storage to data management, data warehousing, and data analytics. Recently, big data application platforms such as Cask Data Application Platform (CDAP)[1] are widely available for developers to more easily design, build, develop, deploy and manage big data applications on top of the existing Hadoop ecosystems. Such platforms significantly reduce the learning time of the developers. However, provided the complex nature of data sources that need to be processed during a disaster situation, a thorough analysis is required to develop a framework for effective use of such platforms.

## Integrated Framework for Disaster Management

A convergence of communication channels, including sensors, mobile apps, social media and telecommunication is important in disaster management applications. As introduced earlier, a big data analytics based application consists of four major processes: collection, preparation, analysis and actions. Here, we analyse the components to develop a framework for our disaster management application and present an ar-chitecture of our application based on this framework. This framework assumes that a cloud infrastructure service with big data tools is available where the application can be developed transparently.

---

[1] http://cask.co/

### *Application Components*

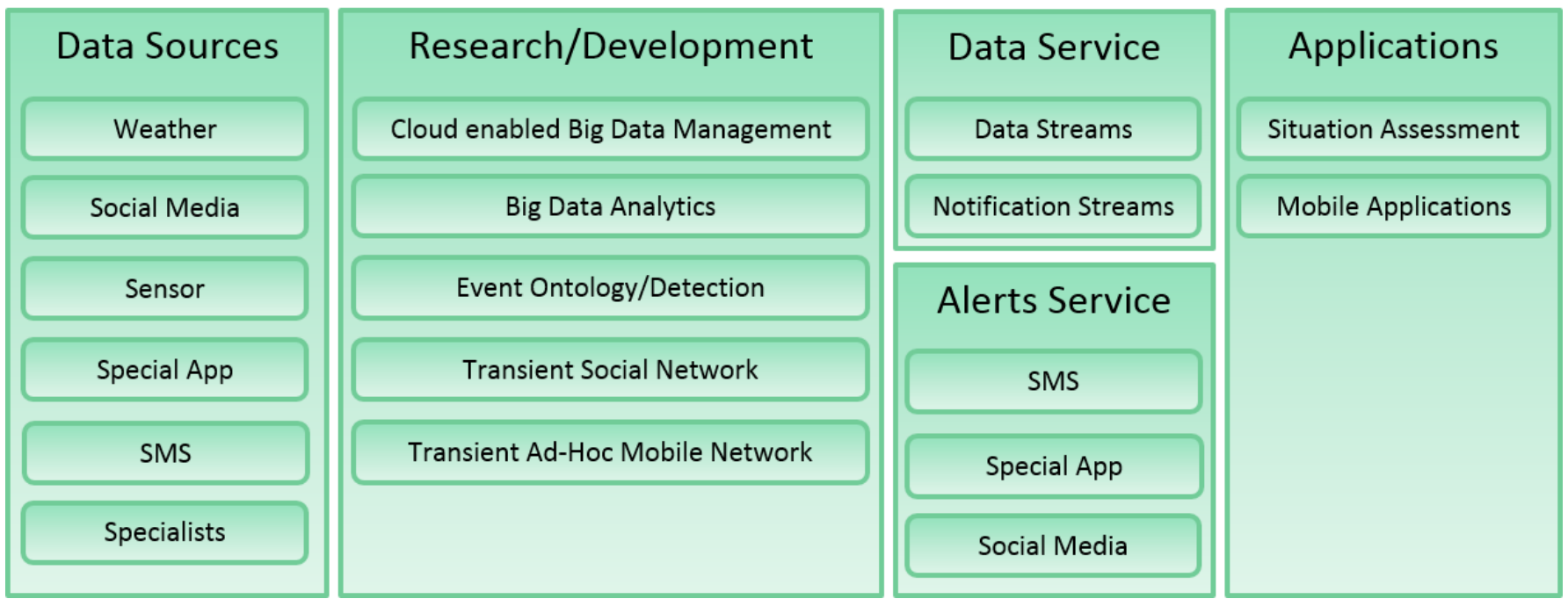

*Figure 5: Components of an Integrated Disaster Application Framework*

An overview of the components required for an integrated disaster application is shown in Figure 5. In this diagram, the *Data Sources* component highlights a list of data sources such as weather forecast, social media, sensors, mobile apps, SMS and specialists. They provide information from different channels that are relevant to disaster management. The development of this component needs to consider the complexity presented due to gathering of data from heterogeneous sources with different volume, velocity and variety. It needs to make the data homogeneous for the downstream analytics. The analytics task is carried out by big data Tools. Event ontology plays a vital role of defining uniform nomenclature of disaster related events that are used for identification of events and generating notifications. During a disaster situation, transient networks such as social and ad-hoc mobile networks enable interactions among people and with the application system which can be used as a platform for taking relevant actions. Finally, the action tasks are carried out by data and alert services. The presentation of the results is performed by application components using web portals and mobile applications.

## *System architecture*

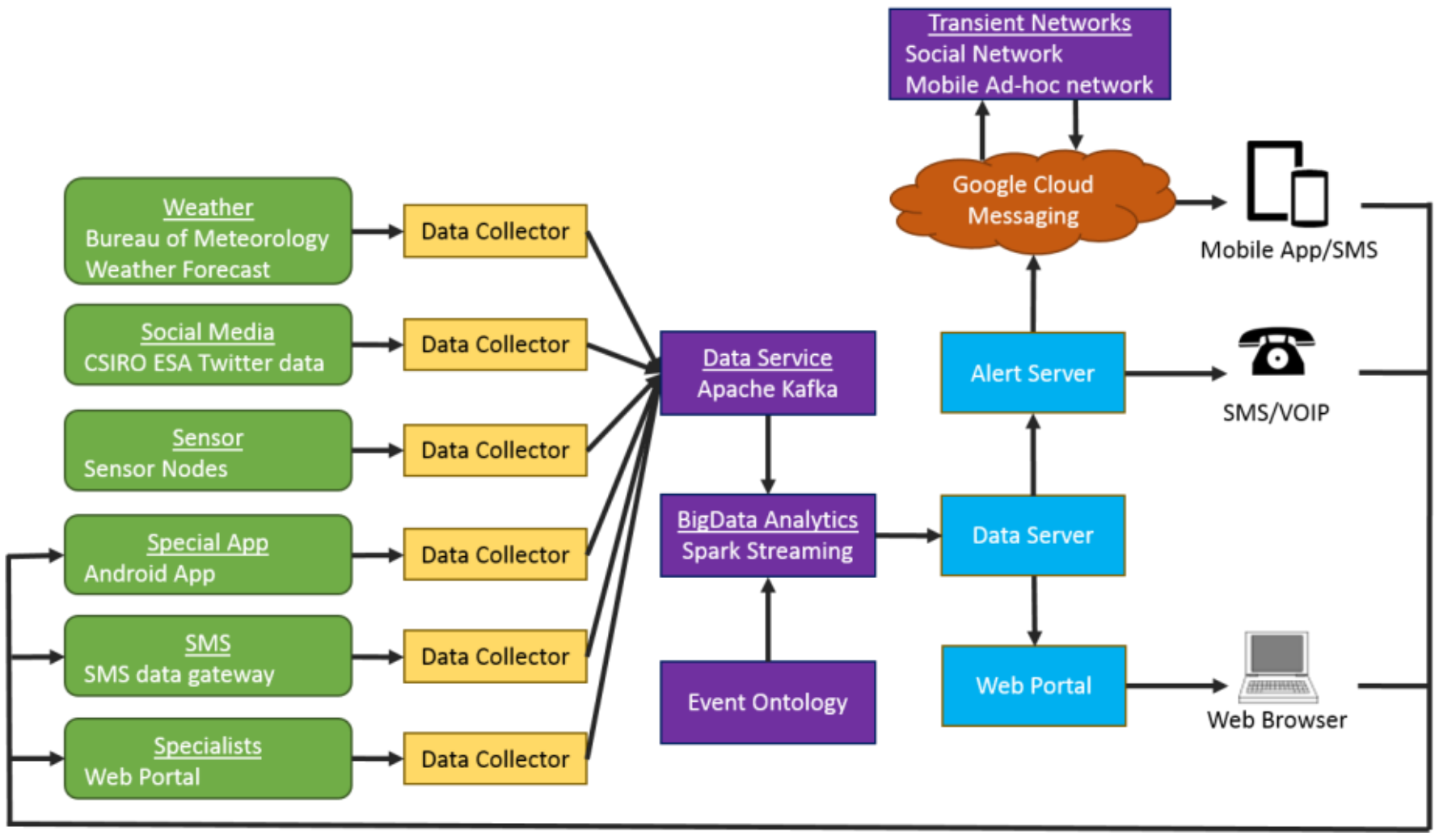

*Figure 6: A system architecture of our disaster management application*

Figure 6 shows our illustrative architecture of the disaster management application using the framework described earlier. In this example, we use the Bureau of Meteorology's (BOM) weather forecast data as weather data source. BOM provides data through their ftp server and updates once twice a day. The data collector communicates with the ftp server to collect the data. The updated forecast data is formatted and sent to the Apache Kafka server. Kafka is a distributed, reliable, and available service that can collect, aggregate, and move large amount of streaming data. Similar to the weather data, data collectors for other data sources such as CSIRO Emergency Situation Awareness (ESA), sensors, mobile apps, SMS gateway and web portal also gather event data and submit the data to Kafka. In addition, event ontology defines disaster conditions which are evaluated by Apache Spark on the event data.

When a disaster event is identified, the event is pushed to event data server. The data from this server is used by the alert server and web portal. The alert server works in an active mode. Based on the subscription for alerts, it sends alerts via the Google Cloud Messaging to android apps, and via SMS to the relevant subscribers. The web portal works in the passive mode. It publishes alerts on a web portal which can be viewed later by the uses. Note that the mobile app, SMS and the web portal can also work as input sources and hence can be designed to both send and receive event messages.

Transient network, especially transient social network in Android apps uses Google Cloud Messaging service to search and join transient networks, and interact with users in the network.

## The Application

In order to demonstrate the application of our proposed application framework for disaster management, we developed a demonstrator as described below showcasing the key components of the architecture.

### *Data sources and their integration*

Effective response to crises and disaster events depends not only on the historical data, but also real-time data from multiple digital channels including social media feeds, text messages from mobile devices and sensor networks. The role of social media (e.g., Twitter, Facebook, etc.) and mobile devices for disaster management has been well studied and documented [20, 35]. Timely analysis of data from these sources can help rescue teams, medics, and relief workers in sending early warning to people, coordinating rescue and medical operations, and reducing the harm to critical national infrastructure.

As outlined in the framework and the architecture, we used data from different sources. From our application's perspective, the data is virtually useless outside the context of disaster. Therefore, we developed mechanisms to initiate data collection and analysis system based on triggers generated by some data sources. Firstly, the weather forecast data is obtained from the Bureau of Meteorology, Australian Government[2] website.

---

[2] http://www.bom.gov.au/

This data is used as the first triggers to start data collection system. Secondly, we established sensor data thresholds to as the second trigger. For this, sensor networks with two nodes, one in Marsfield, NSW and another in Crace, ACT have been deployed. The networks are built using Libelium Waspmote technology, as shown in Figure 7. Besides the internal board temperature and battery status, each node measures temperature, humidity and noise levels. These sensor nodes transmit the measurement data via WiFi and ZigBee network, as well as through the USB ports. Finally, we used CSIRO Emergency Situation Awareness (ESA)[3] as the third trigger to start the data collection system. ESA collects twitter feeds associated with various geographical areas and identifies statistically significant topics related to emergency situations. After authentication, these alerts are accessible through their RESTFul API. A snapshot of the ESA portal is shown in Figure 8.

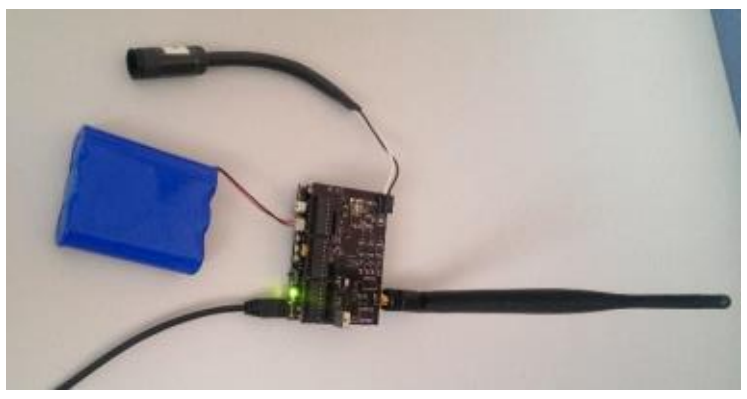

*Figure 7: A sensor node*

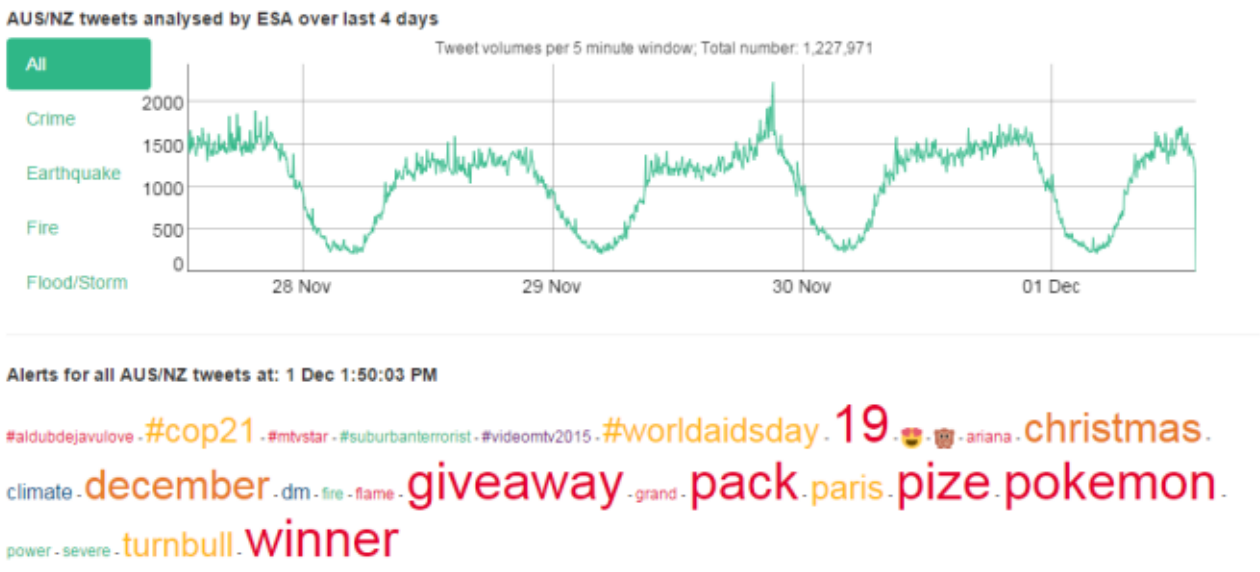

*Figure 8: A snapshot of ESA portal*

---

[3] https://esa.csiro.au/

### *Event Ontologies and Event Detection*

Since big data consist of structured, unstructured, and image data, we needed to develop novel techniques for detecting events. For this, firstly the scenarios of interest were identified to define the entities (i.e., the actors in the scenarios) to develop event ontology. This ontology is able to describe events such as disasters and distress situations and to describe resources and knowledge that are relevant to them. Secondly, for new event detection system, we surveyed several Machine Learning (ML) and Natural Language Processing (NLP) techniques including Naive Bayes, SVM, Random Forest, Logistic Regression or Hidden Markov Models to understand their suitability for detecting events from online streams of data from social media and other sensors [36, 37]. We then developed a novel approach, as described in [38], to detect events in real-time via implementation of clustering and state vector machine learning technique over MapReduce programming framework. The designed event detection model has been implemented as a service in the current prototype system. Furthermore, we created a fully functional system for high performance event detection that includes ready to use virtual machines preconfigured with  NoSQL database (for big data indexing), MapReduce (distributed processing), and Apache Mahout (for event extraction and analytics). Finally, to improve the data organization capacity of HDFS and MapReduce frameworks, we investigated and developed new plugin API that takes into account adjacent data dependency in the data workflow structure. This new modification helps us immensely in improving the performance of MapReduce enabled classifier for event detection from social media, mobile phone, and sensor data including images [39].

### *Mobile App for TSN over DTN*

We developed an Android based mobile app for TSN over DTN allowing users to register, subscribe and send messages as shown in Figure 9.

Firstly, we developed a framework for distributed dissemination of messages over the GSM network by leveraging the capabilities of smartphones in an emergency situation. The fundamental challenge lies in adapting community inspired publish subscribe model in privacy preserving and scalable manner. We developed and implemented a broker-based peer to peer protocol for prioritized delivery of messages in order to support a fast emergency alert and response mechanisms. The proposed framework disseminates messages through streaming channels such as Twitter, Whatsapp, Telegram for social network based interaction. We also implemented a distributed phone book for controlled dissemination of messages along with a framework for handling responses appropriate to alert messages.

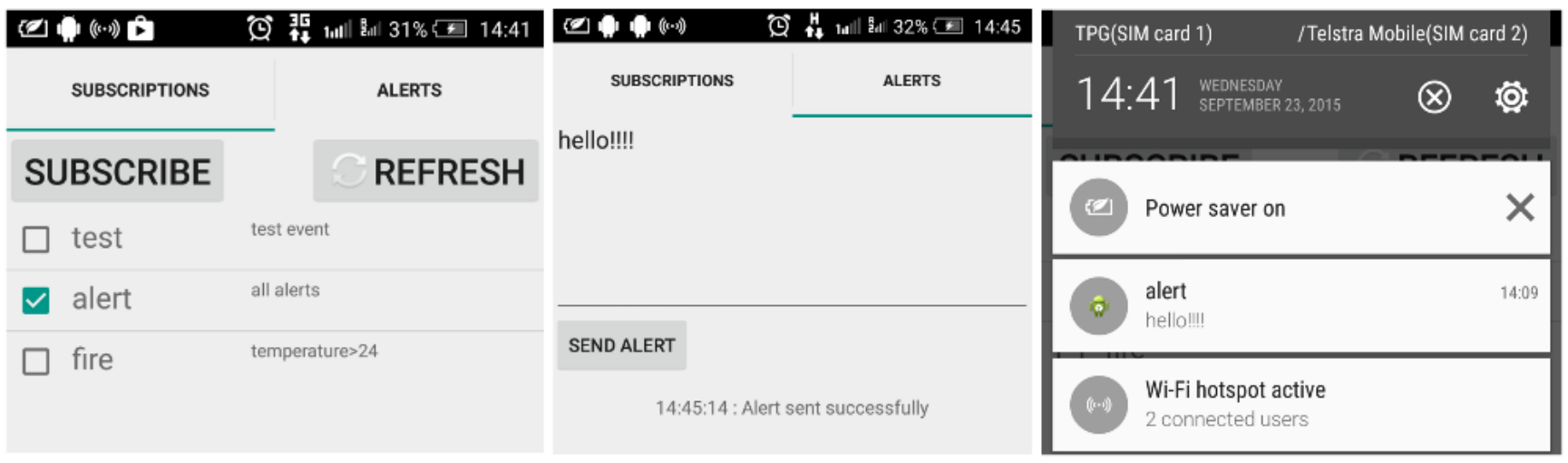

*Figure 9: Android app for alert subscription*

Secondly, in an emergency scenario, the communication goes beyond GSM network. The amount of data generated in emergency situation can overwhelm computer infrastructures not prepared for such data deluge and consequent need for more CPU power. We developed a data-centric application and executed in clouds. The framework handles connections to data sources, data filtering, and utilization of cloud resources including provisioning, load balancing, and scheduling, enabling developers to focus on the application logic and therefore facilitating the development of data-centric emergency applications.

Thirdly, the communication is vital to effective coordination in a disaster response and recovery system. In case of severe disaster, we envisioned a situation where the communication infrastructure is severely damaged and therefore wired communication is not possible. Furthermore, wireless communication depending on infrastructures like GSM communication system may only be inaccessible in patches. Under such situation, establishing a graded response and recovery mechanism appears almost impossible. To solve this problem, we developed a Transient Social Networking (TSN) for disaster situations using opportunistic mesh network of mobile phones. The TSN creation is driven by implementation of a distributed phone book for controlled dissemination of alert messages along with a response mechanism. The proposed system is built by taking into consideration the usual limitations of a smartphone like battery power, processing power, and storage capacity.

Finally, adapting community inspired publish and subscribe model in a privacy preserving and scalable manner is a challenge. We explored some obvious security and privacy issues encountered in a community inspired framework of information dissemination system. An android app has been developed for entire system. The privacy aspects of the message dissemination is handled through message flow labelling techniques.

### *Scalable Cloud Computing*

Emergency management application deployed in the cloud co-exist and share the same infrastructure with other critical applications. Therefore it is important to develop robust resource share estimation for such data-intensive applications. However, developing a reliable resource estimator is quite challenging due to various reasons. We proposed an inclusive framework and related techniques for workload profiling, resource performance profiling, similar job identification, and resource distribution prediction. We developed a Linear Programming model that captures the attributes of virtual machines in cloud environments. The model considers cost, CPU resources, and memory resources of VMs. The virtual machine model was initially designed for multi-objective optimization and was later extended to support multi-criteria optimization in the context of cloud resource selection. The model will be applied in the context of autonomic management of cloud resources to adapt to variations in the application workload [40]. In addition, we have developed a number of algorithms to select, optimise, provision/schedule and monitor cloud resources for different scenarios and resources (refer to [41-55]).

We also developed an evolutionary migration process for application clusters distributed over multiple cloud locations. It clearly identifies the most important criteria relevant to the cloud resource selection problem. Moreover, we developed a multi criteria-based selection algorithm based on Analytic Hierarchy Process (AHP) [56]. Because the solution space grows exponentially, we developed a Genetic Algorithm (GA)-based approach to cope with computational complexities in a cloud market [57].

Hosting of next generation big data applications in domain of disaster management on cloud resources necessitates optimization of such real-time network QoS (Quality of Service) constraints for meeting Service Level Agreements (SLAs). To this end, we developed a real-time QoS aware multi-criteria decision making technique that builds over well-known Analytics Hierarchy Process (AHP) method. The proposed technique is applicable to selecting Infrastructure as a Service (IaaS) cloud offers, and it allows users to define multiple design-time and real-time QoS constraints or requirements. We considered end-to-end QoS optimisation of a typical Streaming Data Analytics Flow (SDAF) that adequately models the data and control flow of an emergency management workflow application. An SDAF consists of three layers: data ingestion, analytics, and storage, each of which is provided by a data processing platform. Despite numerous related studies, we still lack effective resource management techniques across an SDAF. To solve this challenge, we invented a method for designing adaptive controllers tailored to the data ingestion, analytics, and storage layers that continuously detect and self-adapt to workload changes for meeting users' service level objectives. Our experiments, based on a real-world SDAF, show that the proposed control scheme is able to reduce the deviation from desired utilization by up to 48% while improving throughput by up to 55% compared to fixed-gain and quasi-adaptive controllers. Furthermore, we developed a new resource provisioning algorithms for deploying data-intensive applications on hybrid Cloud Computing environments [58]. Here, we have demonstrated its usefulness beyond the disaster situation by deploying Smart-Cities application on Clouds.  We also developed secure Virtual Networking with Azure for hybrid Cloud Computing.

### *Security and Privacy*

We considered security and privacy in both mobile phone apps and sensors based IoT systems. In an Android operating system based mobile phone application, Sinai et al. [59] gives a detailed account of possible attacks on a social navigation. One of the attacks applicable to our TSN over DTN application in a disaster scenario is the Sybil attack. The Sybil attack is an attack wherein a reputation system is subverted by forging identities in peer-to-peer networks. The lack of identity in such networks enables the bots and malicious entities to simulate fake GPS report to influence social navigation systems. The Sybil attack is more critical in a disaster situation where people are willing to help the distressed person. The vulnerability could be misused to compromise people's safety. For example, the malicious user can simulate a fake disaster alarm to motivate a good samaritan to come for help in a lonely place and get harmed. The attacker can also divert the attention of rescue team from the real disaster. Unique Identifier can be used to prevent the Sybil Attack. Hardware identifier such as International Mobile Equipment Identity (IMEI) is more suitable than software identifier which can be easily modified.

Moreover, there is a risk of the privacy breach in case of disclosure of user information. Particularly, users are tried to do their best to communicate with others when they are in distressed situation, e.g. by creating their own TSN. This requires proper information flow models such as Bell-LaPadula model [60, 61], Lattice model [62] and Readers-Writers Flow Model (RWFM) [63] to protect the user's information on the TSN. Combined with proper access control mechanisms, those information flow model can be used to guarantee that the information flow follows the required privacy rules in the TSN and does not leak any critical information to the adversary. For example, in RWFM, the sender can control the readers of a message by specifying their names in the readers list.

Another hard challenge was how to enable end-to-end security and privacy in processing big data streams emitted by geographically distributed mobile phones and sensors. We have investigated a number of techniques for Cloud4BigData application (refer to [64-70] for details). Applications in risk-critical domains such as disaster management need near-real-time stream data processing in large-scale sensor networks. We introduced a new module named as Data Stream Manager (DSM) to perform security verification just before stream processing engine (SPE). DSM works by removing the modified data packets and supplying only original data back to SPE for evaluation. Furthermore, we proposed a Dynamic Key-Length-Based Security Framework (DLSeF) based on a shared key derived from synchronized prime numbers; the key is dynamically updated at short intervals to thwart potential attacks to ensure end-to-end security [64, 71]. DLSeF has been designed based on symmetric key cryptography and dynamic key length to provide more efficient security verification of big sensing data streams. This model is designed by two-dimensional security, that is, not only the dynamic key but also the dynamic length of the key. This model decreases communication and computation overheads because a dynamic key is initialized along with a dynamically allocated key size at both sensors and the DSM without rekeying.

Furthermore, to secure big sensing data streams we have also proposed Selective Encryption (SEEN) method that satisfies the desired multiple levels of confidentiality and data integrity [71]. As the smart sensing devices are always deployed in disaster monitoring area with different sensitive levels, we need to protect data streams based on the level of sensitivity for the efficiency because building a strong encryption channel requires a large computations that consume a lot of battery power. This higher consumption is not acceptable in resource constrained sensing devices. Here, to avoid unnecessary waste of resources, we divided the data streams into three levels i.e. high sensitive, low sensitive and open access and secured them based on their sensitive levels. SEEN can significantly improve the life-time of sensing devices and buffer usage at DSM without compromising the confidentiality and integrity of the data streams.

## Conclusions and Future Work

IoT and cloud enabled BigData applications have potentials to create significant impact in the management of disasters. In this chapter, we firstly introduced the concepts related disaster as well as IoT, cloud and BigData technologies. Specifically, we introduced the concept of using DTNs and TSNs during disaster situations. Secondly, we discussed about a potential disaster situation and how these technologies would work in such a situation. This situation was then used to understand the gaps that required further research and development. Thirdly, we introduced a framework and developed an overall system architecture for disaster management. Finally, we described the details of our prototype system. This prototype system can help understand the components and complexities for development of innovative disaster management applications.

We also identified additional avenues of research in this area. Firstly, our selective encryption method needs to be extended to incorporate information flow model. However, implementation of such system can be challenging. In disaster situation, any security overhead created in mobile phones needs to be minimal in order to preserve the power and elongate the battery life. Our proposed light-weight shared key based encryption method for such applications will be developed in future.

Secondly, we identified the need of developing TSN over an opportunistic mobile mesh network in order to relay data during disaster situation. We have tested preliminary implementation to augment opportunistic network stack for communication over mobile mesh network which consists of nodes with multiple radio interfaces. We will continue to work on an overlay layer as an application to interconnect the islands of disaster hit networks of mobile phones. This application will provide the capabilities of end to end connectivity for TSNs and DTNs. Further research is needed, particularly in the area of developing scheduling plans for enabling hotspot creation while preserving the battery power of the smartphones.

**Acknowledgments** This research is funded by Australia India Strategic Grant AISRF-08140. We also acknowledge the contribution of Rodrigo Calheiros and Aadel Naadjaran Toosi for this work.